\documentclass[12pt]{iopart}
\usepackage{verbatim} 
\usepackage{graphicx}

\newcommand{\ket}[1]{|{#1}\rangle}
\newcommand{\bra}[1]{\langle{#1}|}
\newcommand{\ketbra}[2]{|{#1}\rangle\langle{#2}|}

\begin{document}

\title[
Two interacting atoms in a cavity: exact solutions, entanglement
and decoherence]{
Two interacting atoms in a cavity: exact solutions, entanglement
and decoherence}

\author{J. M. Torres}

\address{Instituto de Ciencias F\'isicas,
Universidad Nacional Aut\'onoma de M\'exico,
C.P. 62210 Cuernavaca, Morelos, M\'exico}

\ead{mau@fis.unam.mx}

\author{E. Sadurn\'i}

\address{Instituto de Ciencias F\'isicas,
Universidad Nacional Aut\'onoma de M\'exico,
C.P. 62210 Cuernavaca, Morelos, M\'exico}

\address{Universit\"at Ulm, Institut f\"ur 
Quantenphysik, D-89069 Ulm, Germany}

\author{T. H. Seligman}
 
\address{Instituto de Ciencias F\'isicas,
Universidad Nacional Aut\'onoma de M\'exico,
C.P. 62210 Cuernavaca, Morelos, M\'exico}

\address{Centro Internacional de Ciencias,
C.P. 62210 Cuernavaca, Morelos, M\'exico}


\begin{abstract}
We address the problem of two interacting
atoms of different species inside
a cavity and find the explicit solutions of the corresponding 
eigenvalues and eigenfunctions using a new invariant.
 This model encompasses various commonly used models.
By way of example we obtain
closed expressions for concurrence and purity as a function of
time for the case where the cavity is prepared in a number state.
We discuss the behaviour of these quantities and and their relative behaviour in
the concurrence-purity plane.
\end{abstract}
 
\maketitle

The system of two two-level atoms (TLA) inside a cavity has attracted considerable
attention, both because it has become experimentally feasible and because
it is the paradigm to study the evolution of entanglement under decoherence.
This combination is remarkable, because entanglement is a central resource 
and decoherence the major impediment for quantum information processing
\cite{nielsenchuang}.
The relation between concurrence and purity of the
central system yields the simplest access to the problem.

Different models of two identical TLA as a central system coupled to a
cavity mode in resonance with the atomic transition as environment
have been studied \cite{xunming08,wang09,aguiar05,tessier03}.
In this paper, we show that one can define a wider
class of such systems that remains solvable in closed form and
includes the above mentioned cases. Specifically we consider
atoms with
different coupling to the cavity mode, different detuning and include
dipole-dipole as well as Ising interactions between the atoms.

We show that the total number of excitations is
 a conserved quantity. Using the basis in which
the corresponding operator is
diagonal, the Hamiltonian will be transformed to block diagonal form, with
maximally $4 \times 4$ blocks.
Interestingly we could use a special case of this solution to construct
an exactly solvable relativistic model \cite{emerson}
with three degrees of freedom,
namely a Dirac oscillator \cite{mosh} coupled to an isospin field.

Note that models with different interacting TLA  and a single excitation
on a continuum of modes have been solved
\cite{pseudomode1,pseudomode2,pseudomode3}
using the {\it pseudomode} approach \cite{pseudomode4}
which in those cases results in a single mode with losses.
While quite similar, the loss term violates the conservation law we use
and thus these are not particular cases of our model.

In order to focus on a particular new aspect, namely the interplay of Ising and
dipole-dipole interaction, we shall choose an example where other features of our
model are simplified.
 Thus  we shall apply the closed solution
to study the time evolution of
concurrence and purity of two interacting TLA with equal
coupling and zero detuning
but arbitrary dipole-dipole and Ising interactions.
The interaction free case basically
provides the borders of the evolution if we look at the interacting
problem in a concurrence-purity
($CP$) diagram, a third boundary being provided by the
relative strength of the two interactions.

Particular cases of the general model, for which solutions are available,
should be experimentally feasible in cavity QED
\cite{kimble06,hemmerich03}.
While dipole-dipole interactions commonly appear in QED, an Ising interaction
might be simulated as proposed in references
\cite{plenio07,porras04,sorensen99}. Wether a particular case, such as the one we discuss, will
actually be measured depends on specific difficulties in forming the initial state, as well as
the amount of interest such a case may arouse.
Some such cases, including initial coherent states, will be studied in a forthcoming paper \cite{forthcoming}. 

Consider the Hamiltonian for two 
TLA coupled to a cavity mode and set $\hbar=1$,
we use the rotating wave approximation and
work in the interaction picture so we end up with
\begin{eqnarray}\label{eq:ham}
H=&\sum_{j=1}^2\left\{
\delta_j\sigma_z^{(j)}+
 g_j\left(a\,\sigma_+^{(j)}+a^\dagger\sigma_-^{(j)}\right)\right\}
+2\kappa\left(\sigma_-^{(1)}\sigma_+^{(2)}+\sigma_+^{(1)}\sigma_-^{(2)}\right)
\nonumber\\&
+J\sigma_z^{(1)}\sigma_z^{(2)}
\end{eqnarray}
where $\delta_j$ is the detuning of the corresponding atomic transition
frequency 
from the frequency of the cavity mode which does not
appear due to our choice of the interaction picture. 

$ g_j$ is 
the coupling to the mode,
$\kappa$ and $J$ are the strengths of the dipole-dipole and 
Ising interactions respectively.
We use the standard 
definitions  of creation and annihilation operators for the harmonic 
oscillator ($a, a^{\dagger}$) and  for the raising
and lowering operators $\sigma_\pm=(\sigma_x\pm i\sigma_y)/2$,
with the Pauli matrices ($\sigma_{x}$, $\sigma_y$, $\sigma_{z}$).

The operator
$I=a^{\dagger} a+1/2\left(\sigma_z^{(1)}+\sigma_z^{(2)}\right)$
provides an additional constant of motion and it 
can be interpreted as 
the number of excitations in the system.
Clearly, $\left[H, I \right]=0 $ and in general 
this is the only commuting observable of 
this problem. Therefore we choose 
the following basis for which $I$ is diagonal
\begin{eqnarray}\label{eq:basis}
\ket{\phi_1^{(n)}}=\ket{n+1}\ket{--}
\quad\quad\quad&
\ket{\phi_2^{(n)}}=\ket{n}\ket{-+}\nonumber\\
\ket{\phi_3^{(n)}}=\ket{n}\ket{+-}
&
\ket{\phi_4^{(n)}}=\ket{n-1}\ket{++}.
\end{eqnarray}
Here $\ket{n}$ describes a state of $n$ photons in the cavity,
$\ket{-}$ and $\ket{+}$ describe the ground and excited states of a TLA
respectively. For any given $n$ they satisfy the relation
$I \ket{\phi_j^{(n)}} = n \ket{\phi_j^{(n)}}$. 
In this basis $H$ is a block-diagonal matrix 
and each block $H^{(n)}$ is a $4\times4$ matrix with
elements 
$\langle \phi_j^{(n)} |  H  | \phi_k^{(n)} \rangle \equiv H_{jk}^{(n)}$. 
Explicitly, one has
\begin{equation}\label{eq:ham44}
H^{(n)}=
\left(
\begin{array}{cccc}
J-\delta_1-\delta_2& g_2\sqrt{n+1}& g_1\sqrt{n+1}&0\\
 g_2\sqrt{n+1}&\delta_2-\delta_1-J&2\kappa& g_1\sqrt{n}\\
 g_1\sqrt{n+1}&2\kappa&\delta_1-\delta_2-J& g_2\sqrt{n}\\
0& g_1\sqrt{n}& g_2\sqrt{n}&J+\delta_1+\delta_2
\end{array}
\right).
\end{equation}
For $n=0$, the basis is reduced to the three states 
$\ket{1}\ket{--}$,
$\ket{0}\ket{-+}$ and $\ket{0}\ket{+-}$. For $n=-1$ it is reduced
to one single $\ket{0}\ket{--}$. This single state 
is stationary and 
represents the situation where both atoms are in the ground state
and there are no photons in the cavity.

Solving the resulting eigenvalue problem implies diagonalizing 
each block of the Hamiltonian.
In general the characteristic polynomial for the eigenvalues
leads to a depressed quartic equation with eigenvalues: 
\begin{equation}
  E_j^{(n)}=\left\{
\begin{array}{l}
-\frac{\sqrt{R+U}}{2}+
\frac{(-1)^j}{2}\sqrt{2 R-U+\frac{Q}{\sqrt{R+U}}},
\quad j=1,2 \\ \\
\frac{\sqrt{R+U}}{2}+
\frac{(-1)^j}{2}\sqrt{2 R-U-\frac{Q}{\sqrt{R+U}}},
\quad j=3,4,
\end{array}\right.
\end{equation}
where we used the following definitions:
\begin{eqnarray}
P=&
\left(\delta_1^2-\delta_2^2+(n+1)\left( g_1^2- g_2^2\right)\right)
\left(\delta_1^2-\delta_2^2+n\left( g_1^2- g_2^2\right)\right)
\nonumber\\
&+J^2\left(
(2n +1)\left( g_1^2+ g_2^2\right)
-2\left(\delta_1^2+\delta_2^2\right)+J^2-4\kappa^2
\right)
\nonumber\\
&+2J\left( g_1^2\delta_1+ g_2^2 \delta_2+
2(2n+1)\kappa g_1 g_2\right)
\nonumber\\&
+4\kappa\left(\delta_1+\delta_2\right) 
\left( g_1 g_2+\kappa\left(\delta_1+\delta_2\right)\right)\nonumber\\
Q=&
4\left(
 g_1^2\delta_2+ g_2^2\delta_1+
4J\left(\kappa^2-\delta_2\delta_1\right)
-2(2n+1)\kappa g_1 g_2
\right)
\nonumber\\
R=&2/3\left((2n+1)\left( g_1^2+ g_2^2\right)+
2\left(\delta_1^2+\delta_2^2+J^2\right)+4\kappa^2\right)
\nonumber\\
S=&
2PR+\frac{Q^2-R^3}{8}
,\quad\quad
T=\frac{4 P}{3}+\frac{R^2}{4}
\nonumber\\
U=&\left(S+\sqrt{S^2-T^3}\right)^{1/3}+\left(S-\sqrt{S^2-T^3}\right)^{1/3}.
\end{eqnarray}
The eigenvectors before normalization read as
\begin{eqnarray}
v_{1,j}^{(n)}=
&\left(E_j^{(n)}-\delta_1-\delta_2-J\right)\times
\nonumber\\&
\left(
\left(E_j^{(n)}+J\right)^2
-n\left( g_1^2+g_2^2\right)
-\left(\delta_1-\delta _2\right)^2
-4\kappa^2
\right)
\nonumber\\
&-2n\left( 
\left(\delta_1+J\right)g_2^2+ \left(\delta_2+J\right)g_1^2
+2\kappa g_1 g_2
\right)
\nonumber\\
v_{2,j}^{(n)}=&\sqrt{n+1}\Bigg( 
2\kappa g_1\left(E_j^{(n)}-\delta_1-\delta_2-J\right)
\nonumber\\
&
+g_2\left(\left(E_j^{(n)}-\delta_1\right)^2+
n\left(g_1^2-g_2^2\right)-\left(\delta_2+J\right)^2\right)
\Bigg)
\nonumber\\
v_{3,j}^{(n)}=&\sqrt{n+1}\Bigg( 
2\kappa g_2\left(E_j^{(n)}-\delta_1-\delta_2-J\right)
\nonumber\\
&
+g_1\left(\left(E_j^{(n)}-\delta_2\right)^2+
n\left(g_2^2-g_1^2\right)-\left(\delta_1+J\right)^2\right)
\Bigg)
\nonumber\\
v_{4,j}^{(n)}=&\sqrt{n(n+1)}
\left(2 g_1 g_2(E_j^{(n)}+J)
+2\kappa\left( g_1^2+ g _2^2\right)\right)
\end{eqnarray}
and the orthogonal transformation which diagonalizes the Hamiltonian is given by
$
V_{j,k}^{(n)}=v_{j,k}^{(n)}/\left({\sum_l v_{l,k}^{(n)\,2}}\right)^{1/2}
$.

By way of example we now treat a special case where we 
calculate the entanglement and purity of the pair of atoms 
considering the cavity mode as environment.
For this purpose it is convenient to start
with product states of cavity and central system functions.
We restrict ourselves to a definite value of the observable $I$ and choose
a number state for the cavity 
{\it i.e.}
\begin{equation}\label{eq:initialstate}
  \ket{\Psi_0}=\ket{n}
  \left(\cos{(\alpha)}\ket{-+}+\sin{(\alpha)}\ket{+-}\right).
\end{equation}
In the same subspace of fixed eigenvalue of $I$, one could also use the state
$\ket{n+1}\ket{--}$ or $\ket{n-1}\ket{++}$ as initial product states.
This type of initial state guarantees that the evolution
stays confined in a four dimensional subspace.
The time evolution of the state vector
under the Hamiltonian (\ref{eq:ham}), can be written as
\begin{equation}\label{eq:tempstate}
  \ket{\Psi(t)}= \sum_{l=1}^4
  B_l^{(n)}(t)\ket{\phi_l^{(n)}}
\end{equation}
with the following coefficients
\begin{equation}\label{eq:tempcoeff}
  B_k^{(n)}=\sum_{k=1}^4
  V_{l,j}^{(n)}
  e^{-iE_j^{(n)}t}
  \left(
  V_{2,j}^{(n)}
  \cos{(\alpha)} 
  +
  V_{3,j}^{(n)}
  \sin{(\alpha)} 
  \right).
\end{equation}
For readability, we shall omit 
the time dependence in the coefficients,
$B_k^{(n)}=B_k^{(n)}(t)$.

Starting from the density matrix of the
whole system $\varrho(t)=\ketbra{\Psi(t)}{\Psi(t)}$, we 
take a partial trace over the cavity degree of freedom to compute
the reduced density matrix of the two TLA, given by
\begin{equation}\label{eq:densmatcavn}
  \rho=\left(
  \begin{array}{cccc}
    |B_1^{(n)}|^2&0&0&0\\
    0&|B_2^{(n)}|^2&\left(B^{(n)}_3\right)^*B_2^{(n)}&0\\
    0&\left(B^{(n)}_2\right)^*B_3^{(n)}&|B_3^{(n)}|^2&0\\
    0&0&0&|B_4^{(n)}|^2
  \end{array}
  \right).
\end{equation}
The purity $P=\Tr\rho^2$ measures the entanglement between
the central system and the environment, {\it i.e.} the decoherence
of the two TLA and we find 
\begin{eqnarray}\label{eq:defpur}
  P&=|B_1^{(n)}|^4+|B_4^{(n)}|^4
  +\left(1-|B_1^{(n)}|^2-|B_4^{(n)}|^2\right)^2.
\end{eqnarray}
The concurrence \cite{wootters98} is used to measure
the entanglement between the atoms. It is defined as
$C(\rho)={\rm Max}\left\{0,\lambda_1-\lambda_2-\lambda_3-\lambda_4\right\}$,
where $\lambda_j$ are the eigenvalues of 
$\left(\rho\,\sigma_y^{(1)}\sigma_y^{(2)}\,\rho^*\,
\sigma_y^{(1)}\sigma_y^{(2)}\right)^{1/2}$ in non-increasing order.
In our case the concurrence is given by
\begin{eqnarray}\label{eq:defconc}
  C(\rho)&={\rm Max}\left\{
  0,\,2|B_2^{(n)}||B_3^{(n)}|-2|B_1^{(n)}||B_4^{(n)}|\right\}.
\end{eqnarray}

Some interesting features can already be 
inferred by inspecting (\ref{eq:defpur}) and 
(\ref{eq:defconc}).
For $n=0$ we have $B_4^{(0)}=0$ and 
the purity has a minimum value of $1/2$. As for the concurrence one can note
the absence of entanglement sudden death \cite{eberly04,eberly06} 
in that particular case.

Now we specialize in the symmetric case with equal couplings to the cavity,
zero detunings, but allow both types of 
interactions between the atoms.  
With these restrictions we are able to find
explicit solutions in the time domain. 
Using the definitions
\begin{eqnarray}
  &\omega_n=\sqrt{4n+2+(\kappa-J)^2}
  \nonumber\\
  &\beta_n=\sqrt{\frac{4n^2+4n}{4n^2+4n+1}}
  \,\,,\quad
  \gamma_n=\frac{6n^2+6n+2}{4n^2+4n+1},
\end{eqnarray}
where $\omega_n$ is a frequency closely related with the 
eigenvalues of the Hamiltonian (\ref{eq:ham}),
and the time-dependent functions
\begin{eqnarray}\label{eq:functions}
  F(t)&=
  \frac{2n+1}{\omega_n^2}
  (1+\sin{(2\alpha}))\sin^2{(\omega_n t)}
  \nonumber\\
  G(t)&=\frac{(\kappa-J)\cos{\left( (J+3\kappa)t\right)}
  \sin{(\omega_n t)}}
  {\omega_n}+
  \nonumber\\
  &
  \sin{\left( (J+3\kappa)t\right)}\cos{(\omega_n t)},
\end{eqnarray}
we find the following solutions for the purity and 
the concurrence as functions of time
\begin{eqnarray}\label{concpurt1}
  C(t)&={\rm Max}\Big\{0,
\sqrt{\left(\sin{(2\alpha)}-F(t)\right)^2
  +\cos^2{(2\alpha)G^2(t)}}
  -\beta_n F(t)
  \Big\}\nonumber\\
  P(t)&=
  1-2F(t)+\gamma_n F^2(t).
\end{eqnarray}

\begin{figure}
  \includegraphics[width=\textwidth]{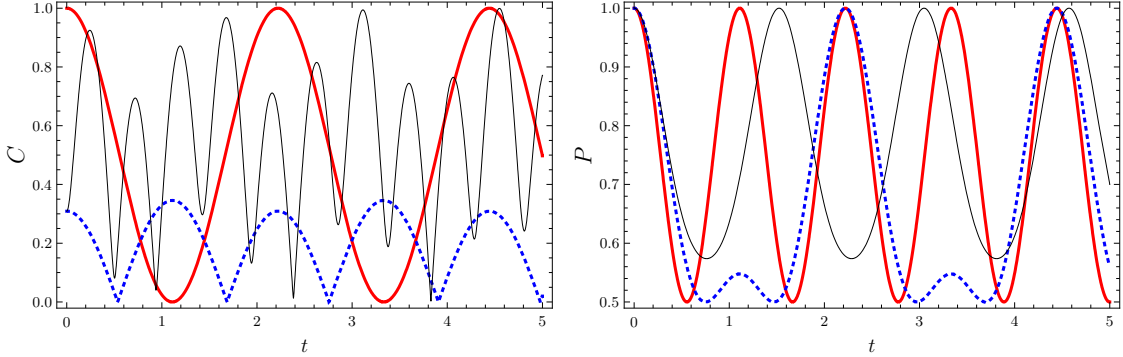}
  \caption{(Colour online)
  Concurrence and purity as function of time 
  for $n=0$ and an initially empty cavity. 
  The red curve corresponds 
  two non-interacting atoms with an initial state
  determined by $\alpha=\pi/4$ (See eq. (\ref{eq:initialstate})) {\it i.e.}
  a maximally entangled pure state.
  The blue dashed curve  shows the behaviour 
  for non-interacting atoms with an initial pure, 
  but not maximally entangled state with $\alpha=\pi/20$.
  In black, the curve for two interacting atoms
  with the same initial state as in the blue dashed curve and
  $\kappa=1.5$ and $J=0$.
  }\label{fig:cpplot0t}
\end{figure}

In figure \ref{fig:cpplot0t} we present these solutions 
for the situation where there are no initial photons in the cavity, namely
$n=0$. We present three cases: in red and blue for noninteracting atoms with
initial states defined by $\alpha=\pi/4$ and $\alpha=\pi/20$ respectively and
an interacting case with $\kappa=1.5$, $J=0$ and $\alpha=\pi/20$ shown in black.
Both quantities, concurrence and purity, 
display oscillatory behaviour, with one frequency in
the noninteracting case and two frequencies in the interacting case as 
can be verified in eqs. (\ref{eq:functions}) and (\ref{concpurt1}).
One can also note that with interaction (black curve)
the concurrence increases while the minimum value of purity is 
greater in contrast
to the corresponding noninteracting case (blue-dashed curve).
Similar behaviour in time domain has already been studied in other references
like \cite{xunming08,wang09,aguiar05,tessier03}. 

\begin{figure}
  \includegraphics[width=\textwidth]{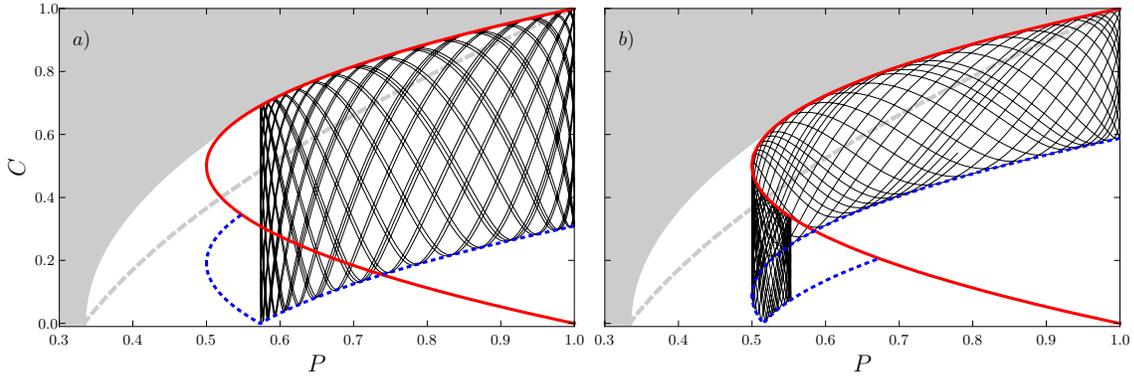}
  \caption{(Colour online)
  $CP-$Plane for $n=0$ and an initially empty cavity. 
  The red curve shows the behaviour in the $CP$-plane for 
  two non-interacting atoms with an initial state
  determined by $\alpha=\pi/4$ (See eq. (\ref{eq:initialstate})) {\it i.e.}
  a maximally entangled pure state.
  The blue dashed curve  shows the behaviour 
  for non-interacting atoms with an initial pure, 
  but not maximally entangled state $a)$ $\alpha=\pi/20$ and $b)$ 
  $\alpha=\pi/10$. In black, the curve for two interacting atoms
  with the same initial state as in the blue dashed curve and
  parametrized by time up to $t=20$. a) $\kappa=1.5$ and $J=0$.
  b) $\kappa=1.5$ and $J=0.87$.
  The gray area indicates 
  $CP$ combinations that can not be obtained in physical states
  and its lower frontier  corresponds
  to the {\it maximally entangled mixed states}. 
  The gray dashed line represents the Werner states.
  }\label{fig:cpplot0}
\end{figure}

The graphs in the time domain look pretty standard and this
does not change if both interactions are present. It is 
therefore convenient
to visualize the joint dynamics 
in a concurrence vs purity plane, the $CP$-Plane.  
Figure \ref{fig:cpplot0} $a)$ shows
the corresponding plane for the curves in figure 
\ref{fig:cpplot0t} with the same colour code, but now the black curve
is parametrized up to $t=20$.
In this plane we have plotted to guide the eye, a gray zone
corresponding to the concurrence and purity combinations 
that can not be obtained 
in physical states and its lower frontier  corresponds to the 
{\it maximally entangled mixed states} (MEMS), 
which for a given value of the purity maximize the 
concurrence \cite{buzek05}.
The gray dashed line is defined by the Werner states
$\rho_W=\xi\frac{I}{4}+(1-\xi)\ket{{\rm Bell}}\bra{{\rm Bell}}$,
$0\le\xi\le1$
\cite{buzek05, pineda06}.

One can note as well, that the dynamic of the interacting (black) case 
is enclosed by the noninteracting curves, the lower bound
given by the blue curve with
the same initial state as the black one, while
the upper bound given by an initial state given by 
$\alpha=\pi/4$.
Perhaps the most important feature here, that 
one can not easily visualize
in the time domain, is that for an initial bell state with $\alpha=\pi/4$
and no interaction between the atoms, the curve (red) follows precisely, as we shall prove below, 
the one that determines the mentioned MEMS.  For this
we need first to obtain the analytic solutions in the $CP$-plane.

We take the explicit solutions in time in eqs. (\ref{concpurt1}), 
with $\kappa=J=0$,
and invert them to find an explicit relation
of the concurrence in terms of the purity. 
In this non-interacting case,
concurrence is represented by up to two different curves in the $CP$-Plane 
\begin{eqnarray}
  C_{\pm}^{(n)}(P;\alpha)&={\rm Max}\left\{0,
  \left|\sin{(2\alpha)}-f_{\pm}^{(n)}(P)\right|
  -\beta_n \,f_{\pm}^{(n)}(P)
  \right\}
\end{eqnarray}
with $\gamma_n$ and $\beta_n$ as given in eq. (\ref{eq:functions})and with
\begin{eqnarray}
  &f_{\pm}^{(n)}(P)=
  \frac{1\pm\sqrt{1+\gamma_n(P-1)}}{\gamma_n}.
\end{eqnarray}
We find two separate cases:
\begin{enumerate}
\item For $
  \frac{1}{2}\arcsin{\left(\frac{n^2+n}{3n^2+3n+1}\right)}<\alpha<\pi/4$, 
  the concurrence in the $CP$-plane is determined by the two curves:
  \begin{eqnarray}\label{eqn:conc1}
  C_{+}^{(n)}(P;\alpha)&,\quad 1-\frac{1}{\gamma_n}\leq P\leq 
  \frac{\gamma_n\left(1+\sin{(2\alpha)}\right)^2}{4}
  -\sin{(2\alpha)}
  \nonumber\\
  C_{-}^{(n)}(P;\alpha)&,\quad 1-\frac{1}{\gamma_n}\leq P\leq 1.
\end{eqnarray}
\item Otherwise, the concurrence is determined only by the curve:
  \begin{equation}\label{eqn:conc2}
  C_{-}^{(n)}(P;\alpha),\quad 
  \frac{\gamma_n\left(1+\sin{(2\alpha)}\right)^2}{4}
  -\sin{(2\alpha)}
  \leq P\leq1.
\end{equation}
\end{enumerate}

\begin{figure}
  \includegraphics[width=\textwidth]{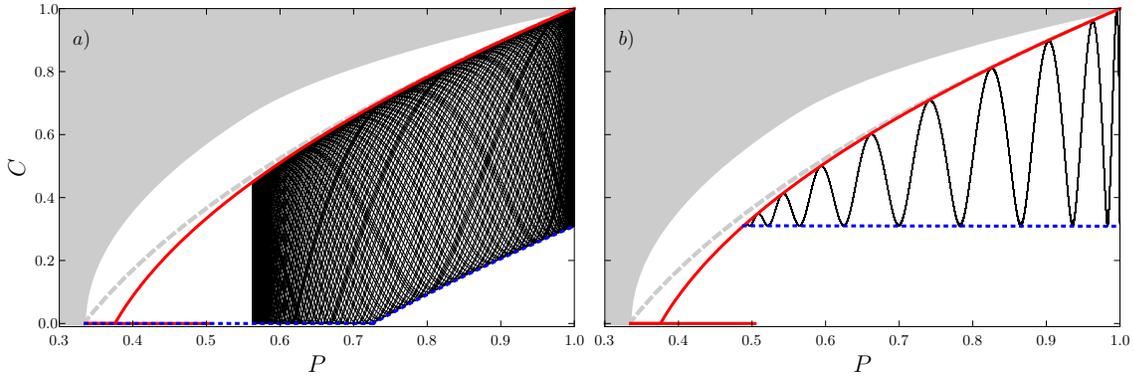}
  \caption{(Colour online)
  $CP-$Plane, same as fig. \ref{fig:cpplot0} but for $n=5$.
  a) $\alpha=\pi/20$ for the black and blue line and
  $\kappa=5.7$ and $J=0.2$ for the black line.
  b) $\alpha=-\pi/20$ for the black and blue line and
  $\kappa=J=5\sqrt{4\times5+2}$ for the black line.
  }\label{fig:cpplot5}
\end{figure}
In figure \ref{fig:cpplot0} 
we show these solutions in the CP-Plane for 
$n=0$  and different values of $\alpha$. 
The red curve shows
the case when the starting state is the symmetric Bell state, $\alpha=\pi/4$.
This solution has the explicit form
\begin{equation}
  C_\pm^{(0)}(P;\pi/4)=\frac{1}{2}\left(1\mp\sqrt{2P-1}\right)
\end{equation}
and it can be seen that in a certain region it 
coincides with the curve for the
MEMS. In fact
$C_-^{(0)}(P;\pi/4)$ coincides precisely with the curve of the MEMS for $5/9\le P\le1$.
The dashed blue curve represents the situation with an initial state determined
by a pure but not fully entangled state.

The behaviour for $n>0$
is qualitatively the same and when one takes as initial state
the symmetric Bell state, $\alpha=\pi/4$,the curves converges 
$(n\to\infty)$ to 
\begin{eqnarray}
  C^{(\infty)}_{-}(P;\pi/4)&=
  {\rm Max}\left\{0,\frac{1}{3}\left(\sqrt{24P-8}-1\right)\right\},
  \quad \frac{1}{3}\le P \le 1
  \nonumber\\
  C^{(\infty)}_{+}(P;\pi/4)&=0, 
  \quad\frac{1}{3}\le P \le \frac{1}{2}.
\end{eqnarray}
Actually in  
figure \ref{fig:cpplot5} we took $n=5$ and the red curve is a very good 
approximation to $C^{(\infty)}_\pm$. 
We note however that in the limit $n\to\infty$ this curve, 
which is actually an upper bound,  lies below the werner curve.
For finite $n$ the curves $C^{(n)}_-(P;\pi/4)$ intersects 
the werner curve in an additional point apart from $C=1$. This
means that there is a small region 
(hardly visible in figure \ref{fig:cpplot5}) above the 
werner curve that can be reached by the dynamics.
We do not write here explicit expressions
for the dashed blue curves, as they can be obtained from equations 
(\ref{eqn:conc1}) and (\ref{eqn:conc2}). 

Figures \ref{fig:cpplot0} and \ref{fig:cpplot5} show in black
the curves for the interacting case for the same initial states as
the blue dashed curves.
One can note 
that in the $CP$-plane the curves now form a 
Lissajous-like figures with their frontier 
defined by the curves $C_\pm^{(n)}(P;\alpha)$ 
for the starting value of $\alpha$ 
(lower frontier) and 
for $\alpha=\pi/4$ (upper frontier). 
Note that 
for increasing values of the difference of the interactions 
the curve in the $CP$-Plane does not fill the 
entire region enclosed by the curves $C_\pm^{(n)}(P;\alpha)$.  
The region filled by the black curve in figure \ref{fig:cpplot0} $b)$
is smaller than in \ref{fig:cpplot0} $a)$, because in \ref{fig:cpplot0} $b)$
we use a larger value of $|\kappa-J|$.
The minimum value of the purity can be calculated as 
$P_{\rm min}={P(t=\pi/2\omega_n)}$, 
this is the lower bound for $P$ for the black curves 
in the figures \ref{fig:cpplot0}$a)$ and 
\ref{fig:cpplot5}$a)$. 
A separate case arises
when 
$\frac{1}{2}\arcsin{\left(\frac{n^2+n}{3n^2+3n+1}\right)}<\alpha<\pi/4$ 
and
\begin{displaymath}
(\kappa-J)^2<
\frac{2 n (n+1)(3\sin{(2\alpha)}-1)+2\sin{(2\alpha)}}{2n+1},
\end{displaymath}
the minimum value is $P_{\rm min}=1-1/\gamma_n$, figure
\ref{fig:cpplot0}$b)$.
If both interactions have the same strength
the curve will fill again the entire area, 
except in the 
case when there are commensurable frequencies in $F$ and $G$, equation 
(\ref{eq:functions}). That is the case of figure 
\ref{fig:cpplot5} $b)$ where the black curve is closed.

We have 
given closed solutions for the dynamics of two different
TLA in a cavity interacting by dipole-dipole and Ising 
interaction.
Many solvable models discussed for two TLA in a cavity
belong to this wider class of exactly solvable models
including a model for a Dirac Oscillator outside the realm of quantum
optics \cite{emerson}.
The effectiveness of the general solution presented was displayed
by calculating the evolution of concurrence and purity
and fully determining the region of its evolution in a $CP$ diagram
in a particular, but interacting case. Interesting features
appear when including both types of interactions.
Intuitively one might think of less decoherence and a more robust
entanglement with increasing interaction between the atoms.
This is true if we have either of the interactions,
but not necessarily if one takes interactions of similar strength.

The parameter space of the model will be further explored in a full length paper.
Interesting situations include placing one TLA outside the cavity or at a
node of the mode and using different detuning as well as coherent or more complicated initial
states.

An interesting perspective would be to extend this technique to situations,
where the couplings are chosen such that -rather than four dimensional spaces
where exact solutions are available- we would have larger but finite spaces
known in molecular physics as polyades, which are accessible to treatments with
Lie algebraic \cite{frank96,iachello91}
and semi-classical \cite{jung99,kellman90} techniques.

Another worthwhile line of research may be to find an even more general class of
solvable models including the one presented here and the ones using a
pseudomode approach  \cite{pseudomode2,pseudomode3}.

Financial support under project \# IN114310
 by PAPIIT, 
Universidad Nacional Aut\'onoma de M\'exico is aknowledged.

\end{document}